# Walk ferroelectric liquid droplets with light

Stefano Marni[1], Giovanni Nava[2], Raouf Barboza[1], Tommaso Bellini[2*], Liana Lucchetti[1*]

1. Dipartimento SIMAU, Universita` Politecnica delle Marche, via Brecce Bianche, 60131 Ancona, Italy. E-mail: l.lucchetti@univpm.it
2. Medical Biotechnology and Translational Medicine Dept., University of Milano, 20054 Segrate, Italy. E-mail: tommaso.bellini@unimi.it

Abstract

We show that the motion of ferroelectric liquid sessile droplets deposited on a ferroelectric lithium niobate substrate can be controlled by a light beam of moderate intensity irradiating the substrate at a distance of several droplet diameters from the droplet itself. The ferroelectric liquid is a nematic liquid crystal in which almost complete polar ordering of the molecular dipoles generates an internal macroscopic polarization locally collinear to the mean molecular long axis. Upon entering the ferroelectric phase droplets are either attracted toward the center of the beam or repelled, depending on the side of the lithium niobate exposed to light irradiation. Moreover, moving the beam results in walking the ferroelectric droplet over long distances on the substrate. We understand this behavior as due to the coupling between the polarization of the ferroelectric droplet and the polarization photoinduced in the irradiated region of the lithium niobate substrate. Indeed, the effect is not observed in the conventional nematic phase, suggesting the crucial role of the ferroelectric liquid crystal polarization.

**Introduction**

The recent discovery of the ferroelectric nematic phase, $N_F$, [1], opened a new chapter not only for the liquid crystal community, but in the whole condensed-matter physics. Beside adding a new, very peculiar, member to the group of ferroelectric materials, the new phase offers a broad range of physical effects to explore, ranging from the behavior of topological defects to surface anchoring [2], response to low frequency electric fields [3] and light, interplay of bound and free electric charges, viscoelastic properties, field-controlled hydrodynamics [3-5], field-order coupling in both the $N_F$ and the pre-transitional regions, behavior in confined geometry [6], just to cite a few examples. In this scenario, we recently performed experiments devoted to characterize the behavior of sessile $N_F$ droplets on ferroelectric solid substrates [4] and found that the combination of fluidity and polarity gives rise to an electromechanical instability induced by the coupling of the LC polarization with that pyroelectrically induced in the solid substrate.

In this work, we instead analyze the effects of the photovoltaic charging of lithium niobate (LN) ferroelectric solid substrates on $N_F$ sessile droplets, at constant temperature. The advantage of manipulating sessile droplets by light are related to the possibility of focusing the beam to small regions of the substrates thus limiting the extent of the charged regions and controlling its distance and position with respect to the droplets, and of quickly reconfiguring it in different illumination geometries. Results show that, under proper experimental conditions, light irradiation gives rise to an instability very similar to the one observed in [4]. Moreover, focused beams impinging on the LN substrate at a certain distance from the sessile $N_F$ droplets allow to optically control their motion. Droplets are attracted toward the center of the illuminated area or repelled away from it depending on which side of the substrate is exposed to light irradiation and can be walked by the light beam over long distances. Our results contribute an additional piece to the collection of intriguing features characterizing the ferroelectric nematic phase and may have potential for future applications.

**Material and methods**

The ferroelectric liquid crystal used in this work is 4-[(4-nitrophenoxy)carbonyl]phenyl2,4-dimethoxybenzoate (RM734). It was synthesized as described in [1] and its structure and phase diagram have already been reported [1, 2, 4]. In this compound the ferroelectric nematic phase appears through a second order phase transition when cooling from the conventional higher temperature nematic (N) phase and exists in the range 133°C < T < 80°C [2,4]. The spontaneous polarization **P** of RM734 is either parallel or antiparallel to the molecular director **n,** defining the average orientation of the molecular axis**,** and exceeds 6 µC/cm$^2$ at the lowest T in the $N_F$ phase [1].

The lithium niobate (LN) ferroelectric substrates are 900 micron thick z-cut crystals. Experiments were performed on iron-doped substrates containing 0.1% mol. of iron with a reduction factor R = 0.02. The bulk spontaneous polarization $P_{LN}$ of LN crystals along the [0001] z-axis is of the order of 70 µC/cm$^2$ and does not depend significantly on T in the explored range since its Curie temperature is much higher (≈ 1140°C). The huge bulk polarization of LN does not however translate in a huge surface charge density because of very efficient compensation mechanisms at the z-cut surfaces, lowering the equilibrium surface charge to only about $10^{-2}$ µC/cm$^2$ [7]. When the crystal is exposed to light with wavelength in the iron absorption spectrum, the surface charge of LN significantly increases because of the photovoltaic effect [8], consisting in the appearance of a photo-induced current according to the scheme $Fe^{2+} + h\nu \rightarrow Fe^{3+} + e^-$. The subsequent charge distribution that takes place inside the crystal gives rise to an internal electric field with saturation values up to $10^7$ V/m, depending on the dopant concentration and on R [8,9]. Before droplet deposition, LN crystals were coated by a layer of fluorolink, as described in [2].

The RM734 droplets used in optical motion control experiments, have an average diameter of 45 µm, as measured with a calibration slide and were obtained starting from bigger droplets as described in the SI. They were deposited on fluorolink-coated LN substrates that were previously slowly heated up to T = 200°C, corresponding to the RM734 isotropic phase. Successively, T was decreased down to 110°C, which is in the ferroelectric range. Noteworthy, the cooling rate was kept slow enough to avoid the droplets electromechanical instability observed in [4] that was triggered by the pyroelectric charging of LN surfaces and required a proper cooling speed.

The light used to induce the photovoltaic effect in LN crystals is a gaussian beam from a frequency doubled Nd:YAG laser ($\lambda$ = 532 nm), with power P in the range (5 -25) mW, focused to a waist w = 35 µm, which corresponds to an intensity ranging from $I = 10^2$ W/cm$^2$ to $I = 5 \times 10^2$ W/cm$^2$. LN substrates were irradiated from below at different distances to the RM734 droplets, holding the temperature fixed at T = 110°C. Polarized optical microscope (POM) observations during light irradiation were carried out and videos of the droplets behavior were recorded with a rate of 25 frames per second.

**Results**

Our first experiment has been devoted to compare the effect of photo-induced charging of LN substrates on RM734 sessile droplets, with the one of pyroelectric charging that we recently reported in [4]. To this purpose, LN was irradiated in correspondence of the droplet position with an unfocused gaussian beam having a diameter slightly larger than that of the droplet itself. As shown in Fig. 1, such a configuration produces an electromechanical instability consisting in the sudden emission of interfacial fluid jets, as observed in [4], indicating that the coupling between the droplet polarization and the one photo-induced in LN has the same features as the coupling with the pyroelectrical LN polarization. In analogy to the interpretation proposed there, we understand this phenomenon as due to the fringing field generated by the photovoltaic charging of LN substrates, which is thus able to affect the $N_F$ droplets behavior. This result opens the way to the possible optical control of ferroelectric LC droplets on LN substrates. Indeed, light is easily controllable: the size and the position of the irradiated region can be varied almost at will and the intensity of the light beam can be tuned in a very short time.

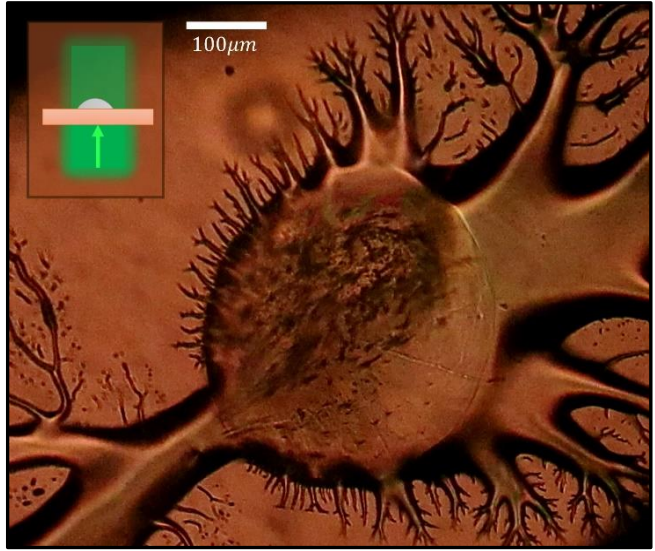

**Figure 1**: RM734 sessile droplet on LN substrate exposed to light illumination. The beam diameter is slightly larger than that of the droplet, as shown in the cartoon on the left-hand side. Shape instability with the emission of interfacial fluid jets that bifurcate and branch, is clearly visible (figure extracted from video S1). $I$ = 20 W/cm$^2$. Droplet diameter 350 µm.

The effect reported in Fig. 1 is independent on the side of the LN substrate contacting the droplet. This is in agreement with the observations in [4] where the sign of the charges of the LN surface that contacts the LC droplet was irrelevant.

The same fringing field generated by photovoltaic charging leads to new effects when the light beam is focused at a distance from the droplets. In this case the droplet retains approximately the same shape but is put in motion by a force generated by light irradiation. Remarkably, the direction of such a force can be either attractive, with droplets that move toward the center of the illuminated region, or repulsive, leading to a droplet motion away from the light spot. The sign of the force depends on the irradiated side of the LN substrate. This is illustrated in Fig. 2, which shows two series of frames extracted from video S2 (Fig. 2b-d) and S3 (Fig. 2b'-d'), available in the SI. As visible, a $N_F$ droplet moves toward the illuminated area or away from it, depending on the direction of the LN bulk polarization with respect to the direction of the incoming light (as in the cartoons on the left-hand side of the figure), which we will refer to as "UP" and "DOWN" in the rest of the paper.

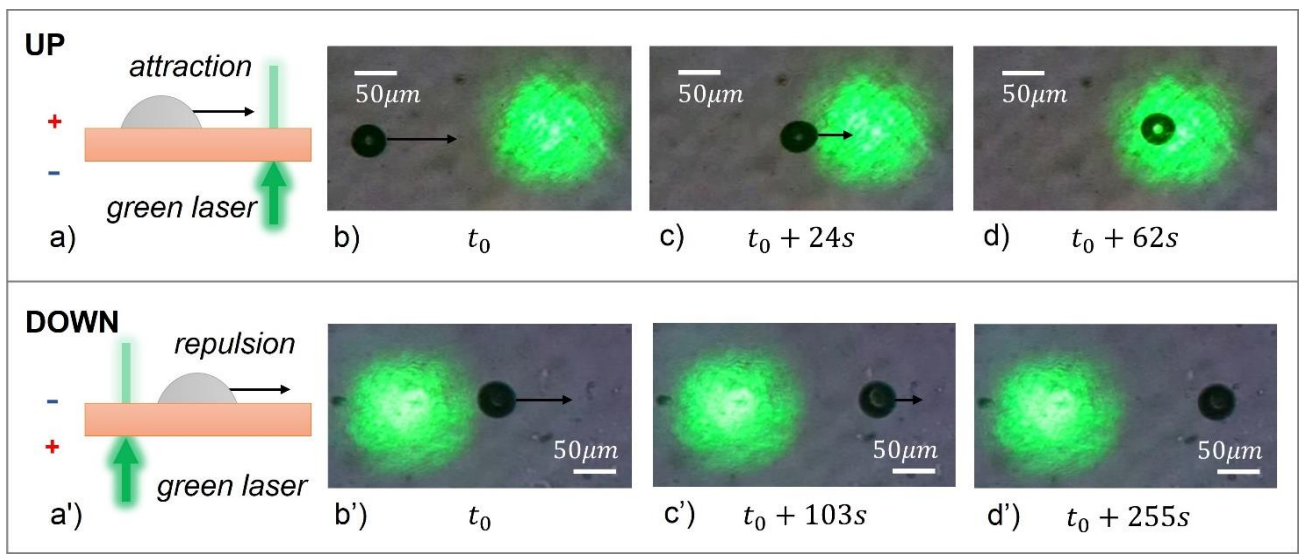

**Figure 2**: a) and a') Sketches of the experimental arrangements in case of droplet attraction a) and repulsion a'): b) and c) video frames taken at different instants showing a RM734 $N_F$ droplet moving toward the center of the illuminate area; b') and c') video frames taken at different instants showing a RM734 $N_F$ droplet moving away from the center of the illuminate area. $I$ = 5 x 10$^2$ W/cm$^2$.

By continuously varying the position of the beam by means of galvanometric mirrors, it is possible to drag the $N_F$ droplet over long distances, as shown in Fig.3 in the case of droplet attraction (frames extracted by video S4). Droplet dragging is also observed in case of repulsion, although with a less controlled trajectory.

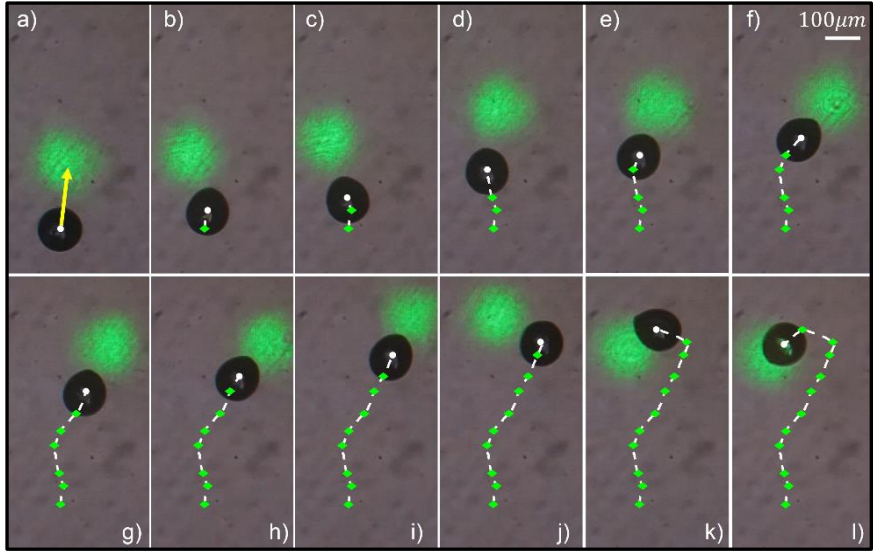

**Figure 3**: Video frames at different instants showing the motion of a RM734 $N_F$ droplet following a Gaussian light spot along a wavy path. Green diamonds indicate the different positions of the droplets through the video and dashed line represents the droplet trajectory. $I$ = 3 x 10² W/cm².

The curves describing droplet motion vs time are reported in Fig. 4 in case of both the UP a) and the DOWN case b). The different curves in each figure refer to different values of the light intensity. Different initial distances between droplet and light spot center have also been chosen to highlight the role of both these parameters. To avoid superpositions and improve the clarity of the two graphs, some of the curves have been translated in time. Remarkably, the curve corresponding to the lowest beam intensity in Fig. 4b (red curve in panel DOWN), shows droplet motion toward the light spot, contrary to what happens at higher intensity.

A close inspection of Fig. 4 reveals that $N_F$ droplets attracted toward the illuminated area stop at the edge of a region corresponding to the beam waist (Fig. 4a); $N_F$ droplets repelled by the illuminated area also stop at a certain distance from its center, that increases with increasing light intensity. This behavior suggests the presence of a pinning force opposing the one due to the interaction with the light beam.

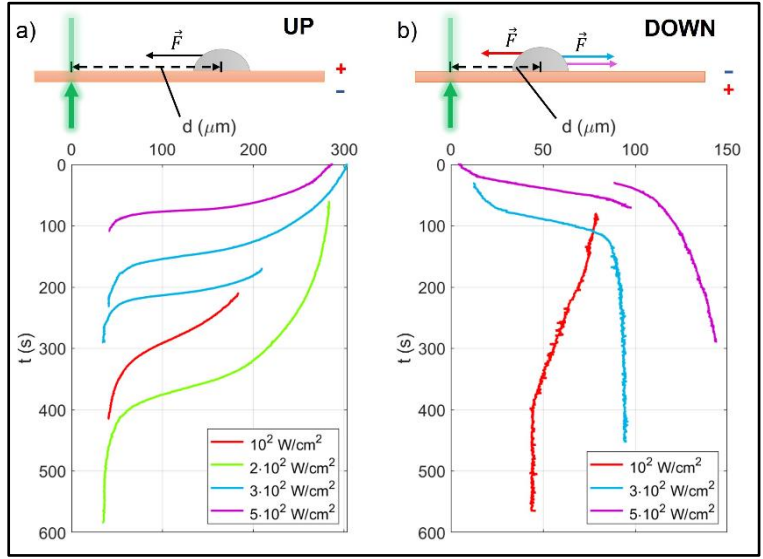

Figure 4: Droplet distance from the center of the illuminated area (d = 0) as a function of time, for the UP a) and DOWN case b). The time t = 0 s corresponds to the beginning of LN illumination, however, to improve the clarity of the figure, some of the curves have been shifted in time so to start at t > 0 s. Different colors indicate different values of the light intensity as reported in the legend; different starting distances have also been chosen. The two cartoons on top help defining the parameters and understanding the experimental arrangement in the two conditions.

From these and similar curves it is possible to extract the time dependence of the droplet velocity *v* for each used value of the light intensity *I*. As an example, Fig. 5 b shows *v* vs *t* in the case of droplet attraction and *I* = 2x $10^2$ W/cm$^2$, for different initial distances from the center of the illuminated area, as extracted from the corresponding *d* vs *t* curves reported in Fig. 5 a. Again, for the sake of clarity, some of the curves are shifted in time. Red dots indicate the initial conditions in terms of initial distance from the light spot center and initial droplet velocity. The three dashed lines identify the values of *d* and *v* corresponding to the same value of *t*. Combining the *v* vs *t* curve with the corresponding *d* vs *t*, one obtains the velocity as a function of the distance from the center of the illuminated area. This is shown in Fig. 5c, where the different curves refer to different initial droplet positions with respect to the center of the light spot. Interestingly, despite a relevant irregularity, the velocity keeps similar values at fixed positions, that is its value is independent on the initial position of the droplet. This indicates that droplets inertia can be neglected in describing the droplet motion, demonstrating that the friction force contains a term dependent on *v*. The average velocity <*v*> is reported in Fig. 5d.

To better understand droplet motion, a friction characterization is necessary. To this aim we performed measurements of droplet motion along a tilted substrate in the absence of light, using a nematic LC of lower viscosity (see SI) to facilitate the experiment. Such measurements showed that friction is indeed composed of two parts: one constant term due to pinning acting as a kinetic friction, and one viscous term proportional to the droplet velocity and due to the internal fluid motion. The equation of motion, neglecting inertia, is thus F = $f_{pinning}$ + $\mu_v v$, where F is the force, attractive or repulsive, due to the action of light and $\mu_v$ is the viscous friction coefficient.

We understand the coupling of the droplet to light as mediated by the fringing field generated by the charge accumulation due to the photovoltaic properties of LN crystals. The simplest form of such coupling is through dielectrophoresis [10]. As a consequence, we expect F to become negligible in the center of the illuminated area (d = 0), where electric field gradients vanish because of symmetry. Therefore, by extrapolating the value of <*v*> for d = 0 (1.1 µm/s in Fig. 5d, red dashed line on the left of the curve), we obtained a relationship between the two friction terms: the pinning force has the value that the viscous friction would have for <*v*> = *v*(d=0). In this way, the force F can be written as F = $\mu_v$ (<*v*> + <*v*(d = 0)>), i.e. F is proportional to *v* through the viscous friction coefficient and thus its dependence on the distance *d* resembles that of <*v*>, as shown in the right part of Fig. 5d. The dashed red line on the right is the power law (F = 18 $d^{(-1.5)}$, with $d$ in µm) that better approximates the force in the range (140 – 320) µm. An estimation of the lower bound of the viscous friction coefficient µ$_v$ has been obtained by performing measurements on tilted LN substrate illuminated so to induce N$_F$ droplet motion toward the top or toward the bottom, as described in the SI. Since within the experimental errors the velocity is the same in the two situations, we could estimate that µ$_v$ is larger than the ratio between twice the component of the gravitational force along the substrate and the uncertainty in the velocity value (see SI). Since we used the lower bound for µ$_v$, the values of the dielectrophoretic force are higher than those in Fig. 5d by a certain amount. We quantified this unknown amount with the number α, in the range 0 < α < 1. The gray area indicates the region within which the force exerted on the droplet is not high enough to overcome friction and generate droplet motion.

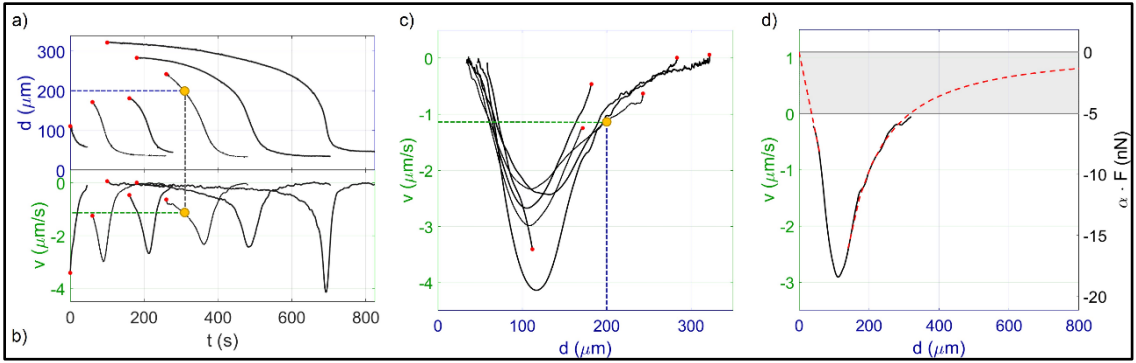

Figure 5: Steps leading to determine the dielectrophoretic force profile as a function of the distance between N$_F$ droplet and light spot center. a) droplet distance from the light spot center for fixed light intensity and different starting positions and b) corresponding droplet velocity as a function of time. Red dots indicate different starting positions while yellow dots identify the values of $d$ and $v$ corresponding to the same time instant; c) droplet velocity as a function of the distance from the light spot center, obtained combining a) and b). Red dots indicate different starting positions while the yellow dot has the coordinate ($d,v$); d) average velocity (left) and dielectrophoretic force (right). The red dashed line on the left is a linear extrapolation of the velocity at $d$ = 0. The one on the right is a best fit of the force for d > 140 µm. The parameter α (0 < α < 1) indicates that the real values of F are higher than those reported in the graph (see text). The gray area corresponds to values of F not enough high to overcome friction and generate droplet motion. $I$ = 2 x 10$^2$ W/cm$^2$.

The average velocities as a function of the droplet distance $d$ from the center of the illuminated area, for all the used values of the light intensity are reported in Fig. 6 for both the UP a) and DOWN case d). Note that the average droplet velocities are negative in a) indicating droplet motion toward the light spot, and positive in d), indicating motion away from the light spot, for all the values of $I$ but the lowest. In this case <v> is negative both in a) and in d). As for the values, at a fixed light intensity, <v> is sensitively higher in case of attractive droplet motion and so is the covered distance. This is also translated in the force profile, which are shown in Fig. 6 b) and e). Both the values and the range are larger in b) than in e), suggesting different values and profiles of the fringing fields responsible for the dielectrophoretic force. In extrapolating the pink and blue curves in Fig. 6 e) for large $d$, we assumed that F would become negative before vanishing at large distances. This is because the electric field has to vanish at large $d$ and thus a range in which dE/d$d$ is negative is required. At the lowest intensity the force becomes positive (red curve) and is analogous to F in Fig. 6b.

To evaluate the electric field profile, we considered the explicit form of the dielectrophoretic force, which is available for the simpler case of spherical shape droplets [10]. In this case, F depends on the gradient of the electric field squared as:

$$F = 2\pi R^3 \varepsilon_m Re\left[\frac{\varepsilon_p - \varepsilon_m}{\varepsilon_p + 2\varepsilon_m}\right]\nabla E^2 \quad (1)$$

where R is the radius of a sphere of complex dielectric permittivity $\varepsilon_p = \varepsilon_p' + i\frac{\sigma_p}{\omega\varepsilon_0}$, dispersed in a medium of complex dielectric permittivity $\varepsilon_m = \varepsilon_m' + i\frac{\sigma_m}{\omega\varepsilon_0}$. σ$_p$ and σ$_m$ are the conductivities of the particle and of the medium, respectively.

When either particle or medium are conductive, at low frequency the imaginary part of the permittivity prevails, and the force becomes [11]:

$$F = 2\pi R^3 \varepsilon_m \left[\frac{\sigma_p - \sigma_m}{\sigma_p + 2\sigma_m}\right] \nabla E^2 \approx 2\pi R^3 \nabla E^2 \qquad (2)$$

Although in a strict sense RM734 is an insulator, the availability of readily displaceable polarization charges makes droplets in the $N_F$ phase behave as conductive droplets since their displaced polarization charges cancel any field internal to the material [4,6]. This is the reason why ferroelectric nematics exhibit an effectively large dielectric coefficient. For this reason, the ratio in square brackets in eq. (2) becomes equal to one and, being $\varepsilon_m$ = 1, the gradient of the field squared can be expressed as the ratio between the force and a term depending on the droplet's volume.

Equation (2) is a rough approximation since (i) the sessile droplet is more similar to a hemisphere than to a sphere and (ii) the droplet is on the top of LN whose dielectric properties are relevant and can locally compensate the surface polarization charges displaced in RM734. We thus argue that eq. (2) overestimates the dielectric force acting in our observations.

Interestingly, the reported optical control of LC droplets has been observed only in the ferroelectric phase. No light-induced droplet movement has been observed in the RM734 N phase nor using conventional nematic liquid crystals, such as 5CB. We understand this behavior as a clear indication that the dielectrophoretic interaction is different in the N and $N_F$ phases. Results suggest that such interaction is stronger when droplets are in the ferroelectric phase. On the contrary, in the nematic phase, the dielectrophoretic force is not strong enough to overcome the pinning force.

The electric field responsible for the dielectrophoretic force can be obtained by integrating the force itself and is reported in Figg. 6 c) (UP) and f) (DOWN), as a function of the distance from the center of the light beam for different values of *I*. A comparison of the two sets of curves shows differences in the field profiles, maximum values, and range. Specifically, both the range and the maximum values are higher in Fig. 6 c), which corresponds to droplet motion toward the center of the light spots. Moreover, maxima are located close to the center of the light spot or at a certain distance, that increases with light intensity, depending on the direction of the droplets motion. The red curve in Fig. 6 f), which corresponds to droplet attraction, is similar in shape to those in Fig. 6 c) but with smaller maximum value and range.

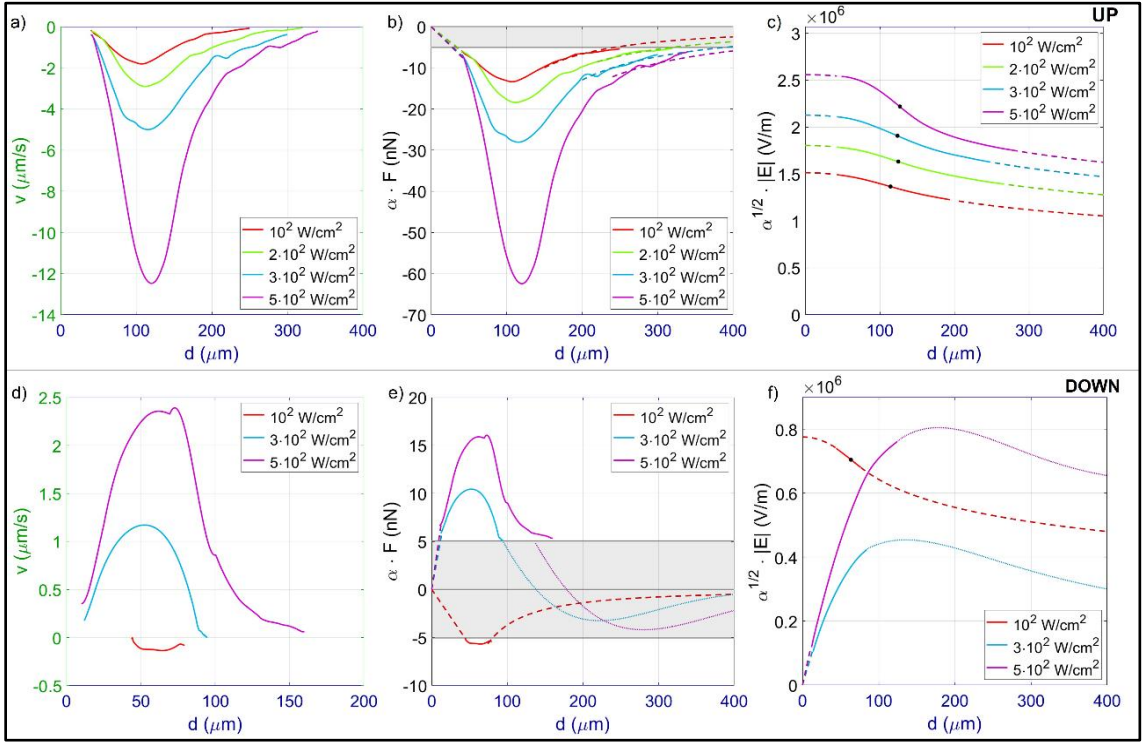

Figure 6: Average droplet velocity, dielectrophoretic force and electric field as a function of the droplet distance d from the light spot center, for the UP (a, b, c) and DOWN (d, e, f) case. Different colors indicate different values of the light intensity. The grey areas in b) and e) are regions of no droplet motion, i.e. regions where the dielectrophoretic force is too weak to overcome friction. In the same panels, dashed lines are power laws that approximate the trend of the forces for values of d outside the actual tracking, while dotted lines are virtual extensions of the real profiles to regions where forces could not be derived from experimental data, due to the absence of motion. Dashed and dotted lines in panels c) and f) represent regions of field values derived from these parts of the dielectrophoretic forces that do not come from experimental data. The black spots in c) and f) mark the inflection points of the curves thus identifying the fields range.

The electric field responsible for the dielectrophoretic forces acting on the $N_F$ droplets, is the fringing field produced by the LN charging due to light irradiation. Although this effect could easily be attributed to the well-known photovoltaic response of LN, the contrasting behavior observed upon inverting the LN crystal, points to a more complex causal chain. A further clue of such complexity is the inconsistency between the expected independence of the LN photovoltaic field on light intensity for values of *I* up to about $10^3 W/cm^2$ [8], and the observation showing a field that clearly depends on *I*.

It should also be noted that the range of the attractive force in the UP case is significantly larger than the laser beam waist (100 vs 35 micron). Among the effects known to take place in LN, the pyroelectric effect is the one most compatible with this observation: in the UP case the range of the force reflects that of local temperature rise due to absorption. The repulsive force in the DOWN case, appears instead to have a shorter range, more compatible with the illuminated area, suggesting that photovoltaic and pyroelectric effect may sum up differently on the two sides. However, such a difference necessarily requires the symmetry between positive and negative charges on the two LN sides to be broken. Any process that would equally modify positive and negative charges, such as a local modulation of the ferroelectric bulk polarization of LN, could not explain our observations. Indeed, the absorbance of the laser green light by the iron-doped LN provides such a symmetry breaking, since the directly irradiated LN side is the location of largest density of iron electron excitation. We reasonably expect such freed electrons to adopt a different spatial distribution when close to the positive vs negative LN surface.

Noteworthy, asymmetry of the two surfaces of z cut LN crystals has already been observed in literature [12-14], both in relation to the efficiency of the surfaces in generating electrostatic fields [13,14], that has been reported to be different for -z and +z surfaces, and in relation to the asymmetric behavior of liquid crystals confined in LN microchannels under light illumination [12]. In this latter case, the combination of photovoltaic and pyroelectric fields was invoked to account for a liquid crystal response dependent on the irradiated surface in an optofluidic chip based on z cut LN crystals.

**Conclusion**

We have demonstrated the all-optical control of sessile ferroelectric liquid droplets deposited on lithium niobate substrates. Droplets are attracted or repelled by the center of the illuminated region depending on the irradiated side of the LN crystal with a range of interaction that is sub millimetric in our conditions, but that may increase by using different surface coatings. Droplets can also be walked by the light beam over long distances. The reported droplet actuation is observed only in the $N_F$ phase, which highlights the crucial role played by the RM734 ferroelectric polarization.

Based on our results, we expect that by reconfiguring the experimental apparatus and structuring light with a Spatial Light Modulator, all the basic droplet handling operations required in a common microfluidic device can be obtained. This opportunity may pave the way to novel technological applications triggered by the peculiar properties of ferroelectric nematics.

We believe that the results reported here contribute an additional piece to the collection of intriguing features characterizing the ferroelectric nematic phase.

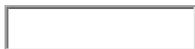